\documentclass[12pt,preprint]{aastex}
\slugcomment{Accepted for publication in the {\it Astrophysical Journal}}
\shortauthors{Mori}
\shorttitle{An X-ray measurement of Titan's atmospheric extent}

\begin{document}
\title{An X-ray measurement of Titan's atmospheric extent from 
its transit of the Crab Nebula}

\author{Koji Mori}
\affil{Department of Astronomy and Astrophysics, Pennsylvania State
University, 525 Davey Laboratory, University Park, PA. 16802, USA}
\email{mori@astro.psu.edu}

\author{Hiroshi Tsunemi and Haruyoshi Katayama\altaffilmark{1}}
\affil{Department of Earth and Space Science, Graduate School of
Science, Osaka University, 1-1 Machikaneyama, Toyonaka, Osaka 560-0043
Japan}

\author{David N.\ Burrows and Gordon P.\ Garmire}
\affil{Department of Astronomy and Astrophysics, Pennsylvania State
University, 525 Davey Laboratory, University Park, PA. 16802, USA}

\and

\author{Albert E.\ Metzger} 
\affil{Jet Propulsion Laboratory, 4800 Oak Grove Drive, Pasadena, CA
91109, USA}

\altaffiltext{1}{present address: National Space Development Agency of
Japan, Tsukuba Space Center, 2-1-1 Sengen, Tsukuba, Ibaraki 305-8505,
Japan}

\begin{abstract}

Saturn's largest satellite, Titan, transited the Crab Nebula on 5
January 2003. We observed this astronomical event with the {\it Chandra}
X-ray Observatory. An ``occultation shadow'' has clearly been detected and
is found to be larger than the diameter of Titan's solid surface. The
difference gives a thickness for Titan's atmosphere of 880 $\pm$ 60 km.
This is the first measurement of Titan's atmospheric extent at 
X-ray wavelengths. The value measured is consistent with or slightly
larger than those estimated from earlier Voyager observations at other
wavelengths. We discuss the possibility of temporal variations in the thickness
of Titan's atmosphere.

\end{abstract}

\keywords {planets and satellites: individual (Titan) --- X-rays:
general}

\section {\label {sec:intro} INTRODUCTION}

Titan is the only satellite in the solar system with a thick atmosphere.
Its atmosphere has a pressure near the surface is about 1.5 times greater than
that of the Earth at sea level and extends much further than that of the
Earth (Coustenis \& Lorenz 1999). Titan's atmosphere is known to
resemble the primitive environment of the Earth's atmosphere in terms of
chemistry, providing us with a laboratory to study the origin of life
(Owen et al.\ 1997).  The atmospheric structure of Titan has been
investigated at radio (Lindal et al.\ 1983), infrared (IR) (Lellouch et
al.\ 1989), optical (Sicardy et al.\ 1990; Hubbard et al.\ 1990), and
ultraviolet (UV) wavelengths (Strobel, Summers, \& Zhu 1992; Smith et
al.\ 1982).  These results came from the Voyager~1 spacecraft during its
encounter with Titan and from a stellar occultation by Titan. Due to the
wavelength dependence of transmissivity, the radio, IR, and optical
observations measure the thermal structure of Titan's atmosphere below
an altitude of 500 km while the UV observation measures it above an
altitude of 1000 km. No direct information has hitherto been obtained at
intermediate altitudes where the X-ray observation is effective.

On 5 January 2003, the Saturnian system passed across the $2'$ wide
X-ray bright region of the Crab Nebula. The Saturnian system has a
conjunction with the Crab Nebula every 30 years. However, because of
an average offset of a few degrees, it rarely transits the Crab
Nebula. Although a similar conjunction occurred once in January of
1296, the Crab Nebula, which is a remnant of SN1054, must have been
too small to be occulted. Therefore, this may be the first transit
since the birth of the Crab Nebula. The next similar conjunction will
take place in August of 2267, making its occurrence in 2003 a
``once-in-a-lifetime'' event.  The Saturnian system is a million times
brighter in visible light than the Crab Nebula whereas the Crab Nebula
is a million times brighter in X-rays than the Saturnian system. This
prevented us from observing the transit in the optical, but provided a
unique opportunity for an X-ray observation which had never been
performed. The Crab Nebula is one of the brightest synchrotron sources
in the sky and, thus, makes an ideal diffuse background light source
to study X-ray shadows of interesting objects.

Here, we report results from an observation of this historical event
with the Advanced CCD Imaging Spectrometer (ACIS) aboard the {\it
Chandra} X-ray Observatory. {\it Chandra} has an angular resolution of
$0.^{\!\!\prime\prime}5$ and can resolve Saturn, its rings, and the
satellite Titan, whose angular size is about $1''$. Unfortunately,
it was not possible to observe the transit of Saturn
due to {\it Chandra}'s concurrent passage through the Earth's radiation
zone. Only Titan was observable because {\it Chandra} had passed through
the radiation zone by the time Titan transited 12 hours later. We
describe the observation in \S \ref{sec:obs}. The data analysis and
results are presented in \S \ref{sec:ana}. Comparison with previous
observations and implications of this observation will be discussed in \S
\ref{sec:discussion}.

\section {\label{sec:obs} OBSERVATION}

\begin{figure}
\epsscale{1.0}
\plotone{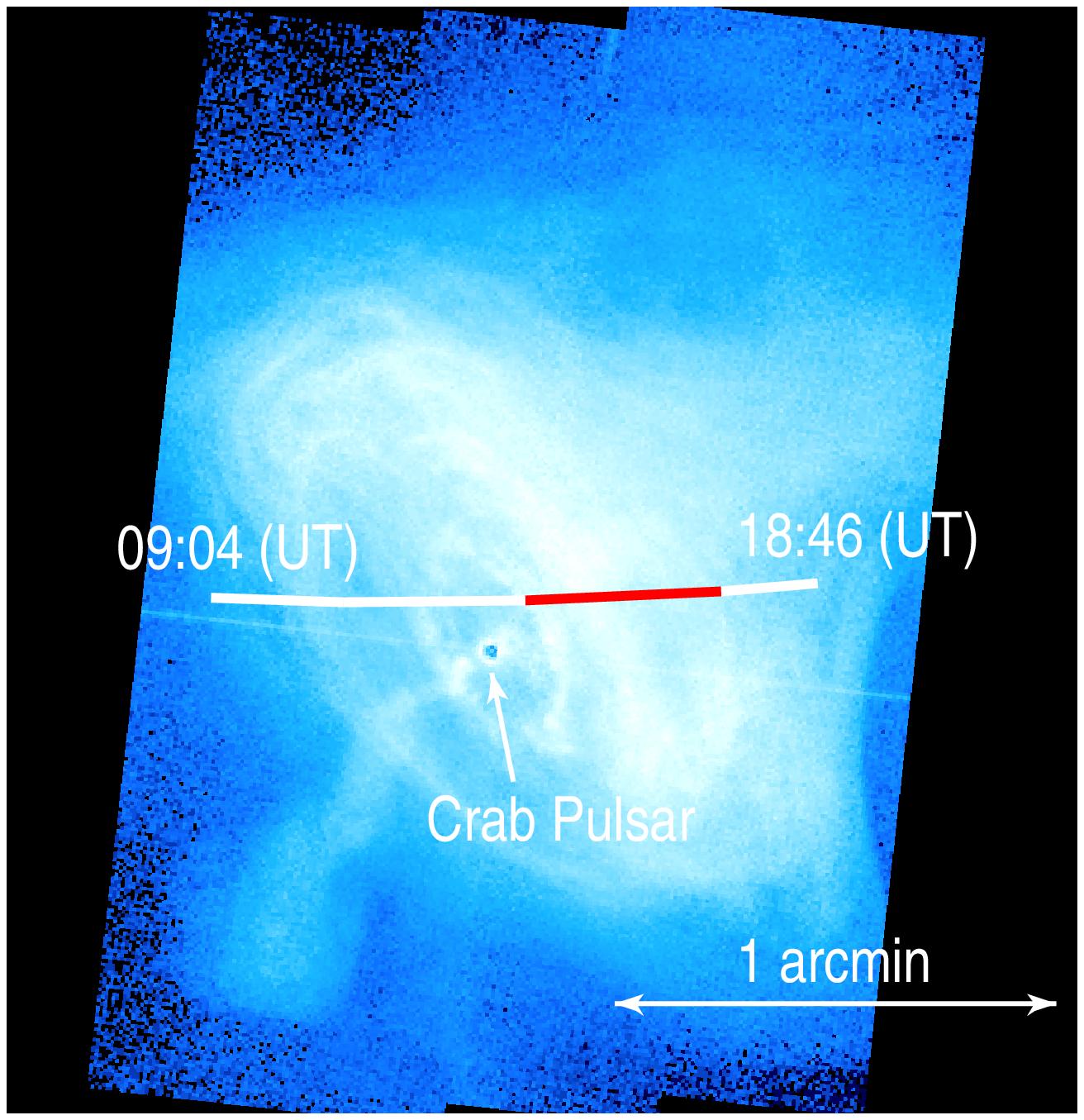}
\caption{
{\it Chandra} image in equatorial coordinates with intensity
displayed on a logarithmic scale. The solid curve running from east to
west represents the Titan transit path seen from {\it Chandra}. Red
and white colors show the assigned ``high flux time interval'' and ``low
flux time interval'', respectively (see text). The observation started at
09:04 and ended at 18:46 (UT) on 5 January 2003. The right bottom
arrow defines one arc-minute of scale. A hole at the pulsar and a
narrow line through the pulsar are instrumental effects.
\label{fig:sky}}
\end{figure}

The observation was performed from 09:04 to 18:46 (UT) on 5 January
2003. Figure~\ref{fig:sky} shows the Titan transit path on the Crab
Nebula. In order to avoid event pile-up and telemetry saturation,
which made difficulties for analysis of previous {\it Chandra}
observations of the Crab Nebula (Weisskopf et al. 2000; Hester et
al. 2002), we inserted a transmission grating, shortened the CCD frame
time from the nominal value of 3.2 seconds to 0.3 seconds, and adopted a
small subarray window ($50'' \times 150''$). These observational modes
did work efficiently to result in no telemetry saturation and no
apparent event pile-up. We will carefully investigate the possibility of
the event pile-up effect on our result in \S \ref{sec:ana3}. Since the
restricted window could not cover the whole Titan transit path, we
changed the pointing direction twice during the observation. In spite of the
time loss due to the two maneuvers, we obtained an effective exposure
time of 32279 seconds, which corresponds to 92\% of the total observing time
duration.

\section {\label{sec:ana} ANALYSIS AND RESULTS}

\subsection {Reprojection To the Titan Fixed Frame}

\begin{figure}
\epsscale{1.0}
\plotone{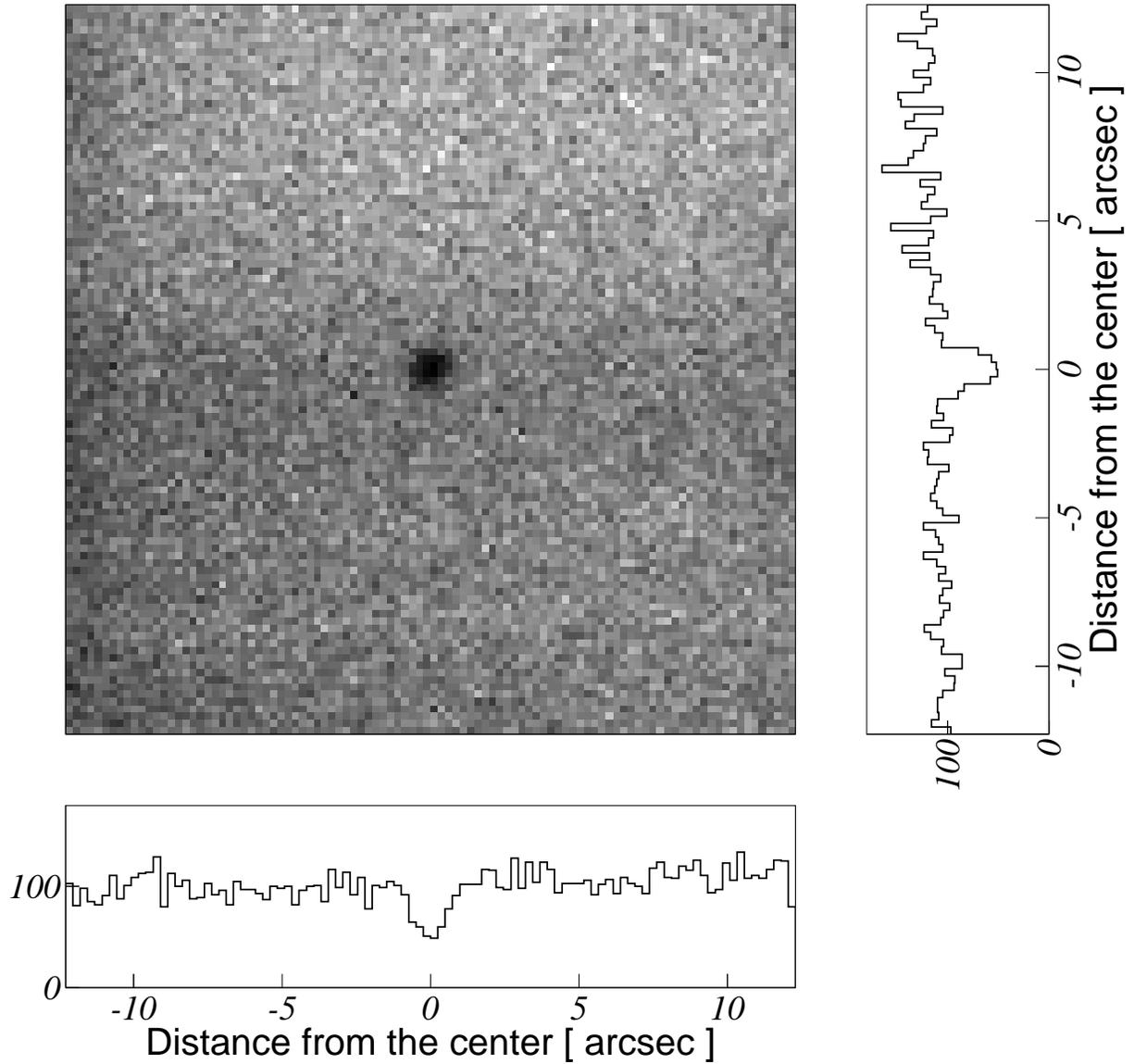}
\caption{
{\it Chandra} image in the frame fixed with respect to Titan.
Bottom and right histograms show the density profiles (counts
bin$^{-1}$) along a horizontal and vertical line through the shadow,
respectively. The image was made with $0.^{\!\!\prime\prime}246$ pixels
to enhance the detection of the shadow. The image size is $25'' \times
25''$.
\label{fig:shadow}}
\end{figure}

No hint of the Titan transit can be seen in the standard sky image in
Figure~\ref{fig:sky}. Titan moved too fast to cast an observable shadow against
the Crab Nebula in the sky image. In order to search for a shadow, we reprojected each photon's
position to a frame fixed with respect to Titan.  We used
CIAO\footnote{{\it Chandra} Interactive Analysis of Observations. See
http://cxc.harvard.edu/ciao/} tool {\tt sso\_freeze} for the
reprojection. Before applying {\tt sso\_freeze}, we removed the
pixel randomization, which was applied by default in the standard data
processing, and applied our subpixel resolution method (Tsunemi et al.\ 2001;
Mori et al.\ 2001)\footnote{See
http://asc.harvard.edu/cont-soft/software/subpixel\_resolution.1.4.html}
to obtain the best available spatial resolution.
Figure~\ref{fig:shadow} shows the {\it Titan fixed frame} image. A clear
shadow can be seen, which was made by Titan's occultation of the Crab
Nebula. Measurement of the effective radius of the occultation shadow
provides us with information on Titan's atmospheric extent.

\subsection {Effective Radius of the Occultation Shadow}

Figure~\ref{fig:data_profile} shows a radial profile of the photon
number density from the center of the occultation shadow (black
points). From this radial profile, we computed the effective radius of
the occultation shadow, $R_{shadow}$. 

\subsubsection{\label{sec:ana1} Method}

We assumed that the distribution of photons in
Figure~\ref{fig:shadow} is symmetric with respect to the center of the
occultation shadow. The {\it observed} photon number density as a
function of the distance from the center of the occultation shadow,
$D_{obs}(r)$, can be given by a convolution of the {\it intrinsic}
photon number density, $D_{int}(r)$, and the point spread function (PSF),
$PSF(r)$:

\begin{equation}
D_{obs}(r) = \int D_{int}(r') PSF(|{\bf r-r'}|) d{\bf r'}.
\label{eqn:density}
\end{equation}

\begin{figure}
\epsscale{1.0}
\plotone{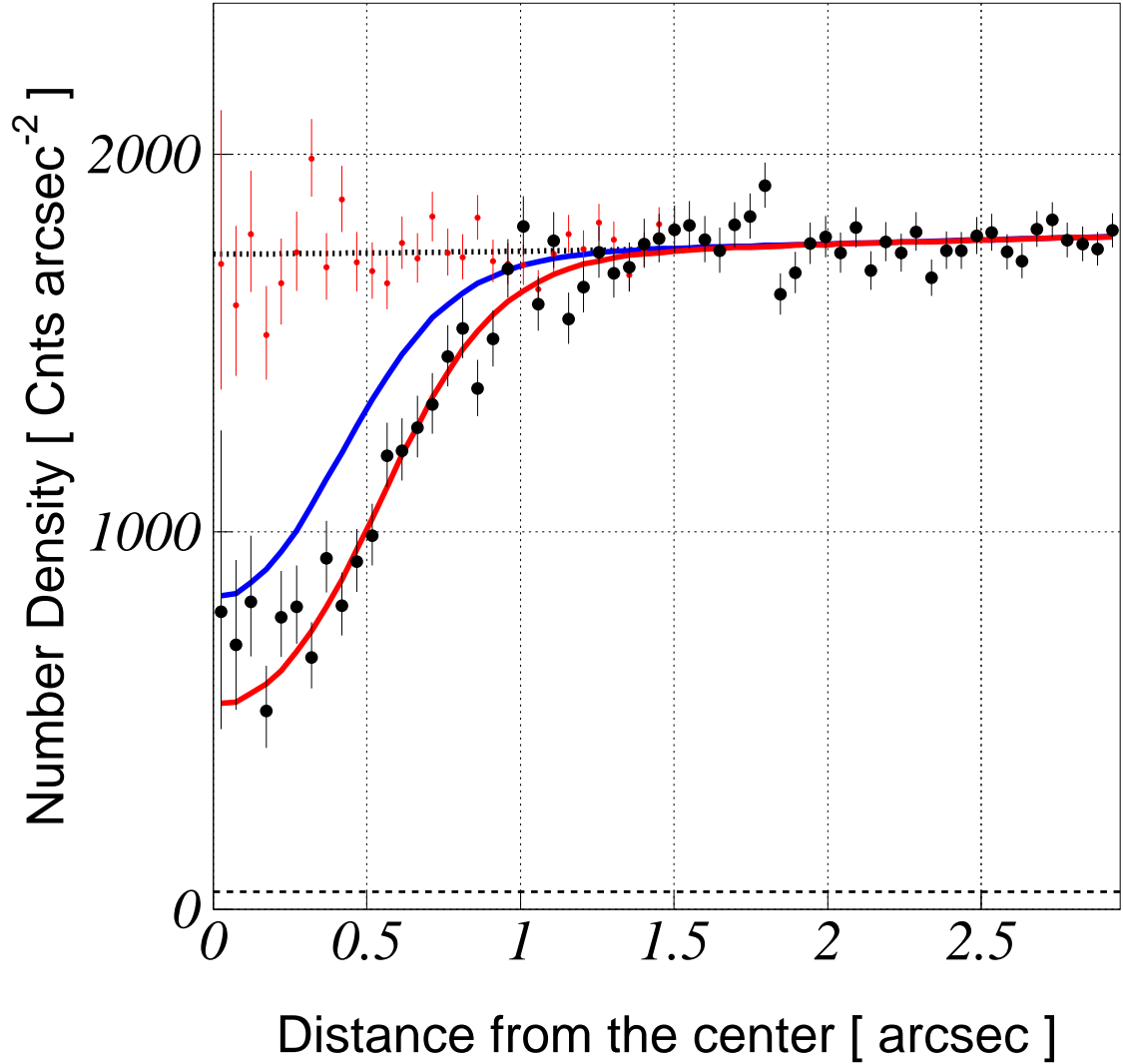}
\caption{
The radial profile of the photon number density from the center of the
occultation shadow (black points). The red curve shows the best fit of
the disk model simulation to the radial profile data, while the blue
curve shows the simulated shadow profile of Titan's solid surface. The
red points represent the background across the shadow. The dotted and
dashed lines indicate the slope of the background and the contribution
of the trailing events, respectively. The dotted line was determined
based on the red points within the shadow and black points outside of
it.
\label{fig:data_profile}}
\end{figure}

\noindent 
The {\it Chandra} PSF has a very sharply peaked core with extended wings
and is strongly energy dependent due to larger scattering of higher
energy photon by the mirror surface (Chandra Proposers' Observatory
Guide 2002). In general, it is quite difficult to obtain a precise
analytical form of the PSF, which prevents us from solving
equation~\ref{eqn:density} to obtain $D_{int}$ directly from $D_{obs}$.
Instead, its numerical form can be generated through a raytrace code
which has been determined by the {\it Chandra} team on the basis of both
ground-based and on-orbit calibrations. Therefore, we performed Monte
Carlo simulations incorporating the numerical form of the PSF. We
approximated $D_{int}$ by a step function with a threshold radius,
$R_{disk}$. We fitted the data points in Figure~\ref{fig:data_profile}
with $D_{obs}(r)$, varying $R_{disk}$ to obtain the best fit.
$R_{shadow}$ is defined by $R_{disk}$ which gives the minimum
$\chi^{2}$. Justification for the approximation using the step function
will be discussed in \S \ref{sec:discussion}.

The PSF was derived from raytrace tools ChaRT\footnote{{\it Chandra} Ray
Tracer. See http://asc.harvard.edu/chart/} and MARX\footnote{See
http://space.mit.edu/CXC/MARX/}. ChaRT provides the best available
mirror response at any off-axis angle and for any spectrum.  MARX reads
the output of ChaRT and creates the PSF taking into account the detector
responses and aspect reconstruction uncertainties. We obtained the PSF
appropriate for the observation conditions: at off-axis angle of about
$45''$ and for the observed spectrum. The output PSF from MARX includes
the pixel randomization. Since the pixel randomization broadens PSF, we
sharpened it by the appropriate amount. To account for our application
of the subpixel method, we further sharpened the PSF by 5\% in terms of
the Half Power Diameter (HPD), which is the expected improvement for
this off-axis angle on the ACIS-S CCD (Tsunemi et al.\ 2002; Mori et
al.\ 2002). The HPD of the resultant PSF was $0.^{\!\!\prime\prime}844$.
We will discuss the validity of this PSF in \S \ref{sec:ana3}.

In order to compare the results of the Monte Carlo simulation with the
data, we must also account for gradients in the sky background level and
for instrumental effects like trailing events. A big advantage of our
observation was the ability to determine the background level across the
shadow based on knowledge of the surface brightness of the Crab Nebula
along the Titan transit path. We derived the background profile across
the shadow by averaging two profiles which were taken centered at $3''$
ahead of and $3''$ behind of the occultation shadow center. The
background profiles are presented as red points in
Figure~\ref{fig:data_profile}. Then, the gradient of the background was
determined by fitting the red points within the shadow ($r<
1.^{\!\!\prime\prime}5$) and the black points outside of it with a
quadratic function, which is shown with a dotted line in
Figure~\ref{fig:data_profile}.

The trailing events are defined as events detected during the 
CCD readout. Since the positions of the trailing events are recorded
improperly along the readout direction, they cause an offset in the
image. The contribution of the trailing events to the total photon number
density can be evaluated because its level is proportional to the
fraction of the total effective exposure time spent in transferring the charge
across the X-ray bright region, $F_{trailing}$. In the case of ACIS, it takes
40 $\mu$sec to transfer the charge from one pixel to another. The width
of the X-ray bright region along the readout direction is about 210 pixels
($\approx 103''$; see Fig.~\ref{fig:sky}). Therefore, it takes 
0.0084 seconds to transfer the charge
across the X-ray bright region. Since CCD frame time is 0.3 seconds,
$F_{trailing}$ becomes $0.0084/(0.0084+0.3)\approx 2.7$\%. The resultant
estimate for the background level due to the trailing events is shown as a dashed line in
Figure~\ref{fig:data_profile}.

\subsubsection{Result}

The first 30 data points from the center ($r< 1.^{\!\!\prime\prime}5$)
were used for the $\chi^{2}$ test because data points further from the
center hardly affected the result. Figure~\ref{fig:chi2} shows
$\chi^{2}$ as a function of $R_{disk}$. The $\chi^{2}$ values are well
described with a quadratic function. The best fit value was obtained
as $R_{shadow} = 0.588 \pm 0.011$ arcseconds with $\chi^{2}$\,/\,d.o.f
of 30.36\,/\,29. The uncertainty is 68.3\% (1$\sigma$) confidence
level. The best fit curve is shown in Figure~\ref{fig:data_profile}
(red line) along with a curve simulating $R_{shadow} =
0.^{\!\!\prime\prime}437$ (blue line) which corresponds to the radius
of the Titan solid surface (2575 km). The difference between those two
curves is attributed to X-ray absorption by Titan's atmosphere.

\subsubsection{ \label{sec:ana3} Systematic checks}

The validity of the PSF, the effect of event pile-up, and the
contribution from the intrinsic background of ACIS were examined. The
accuracy of the adopted PSF was checked using a point
source in the Orion Nebula Cluster, CXOONC~J053514.0-052338, located
at a J2000 position of $\alpha=$ 05$^{\mathrm h}$35$^{\mathrm
m}$14\fs06, $\delta=-$05\arcdeg23\arcmin38\farcs4 (Getman et al.\ in
preparation). The point source has an off-axis angle of $45''$,
comparable with that of Titan in our observation. We processed the
data of the point source in the same manner described in \S
\ref{sec:ana1}; removing the pixel randomization and applying the
subpixel resolution method. The PSF was obtained again through
ChaRT/MARX, but appropriate for the spectrum of the point source, and
was sharpened again in the same manner described in \S \ref{sec:ana1}.
We confirmed that the PSF originally obtained from ChaRT/MARX had a
HPD 15\% larger than that of the point source, but that the PSF
sharpened for our data processing effects was consistent with the
point source data to within a few percent. Next, we studied the
dependency of $R_{shadow}$ on the PSF by performing simulations using
PSFs with 10\% larger and smaller HPDs than that of adopted one.  In
an ideal case, even if a wrong PSF was used, $R_{shadow}$ would be
always the same although the $\chi^{2}$ value would increase and the
fit would not be statistically acceptable. In reality, $R_{shadow}$
depended on the HPD; $R_{shadow}$ became larger and smaller by about
0.018 arcseconds which is comparable to about a 1.6 $\sigma$
statistical error.  However, considering that the adopted PSF is
accurate to a few percent, we can safely assume that this dependency
is negligible compared with the statistical error. We note that those
simulations using broader and narrower PSFs resulted in higher
$\chi^{2}$ value by about 3 than the simulation described in \S
\ref{sec:ana1}, also supporting the validity of the adopted
PSF. Therefore, we conclude that the adopted PSF is reliable and its
uncertainty hardly affected the result.

In order to examine the pile-up effect, the data were divided into two
time intervals according to flux level, ``high flux time interval''
and ``low flux time interval''. The division is indicated in
Figure~\ref{fig:sky}. The average background count rates in the high
and low flux time intervals are 8.1 and 4.0 $\times 10^{-2}$ counts
sec$^{-1}$ arcsec$^{-2}$, respectively.  Separate calculations of
$R_{shadow}$ for the two data sets were performed in the same way
described above. Plots of $\chi^{2}$ as a function of $R_{disk}$ are
shown in Figure~\ref{fig:chi2}. The resultant values were $R_{shadow}=
0.599 \pm 0.016$ and $0.581 \pm 0.016$ arcseconds for the high and low
flux time intervals, respectively. Since the values are statistically
consistent with each other and with the result obtained using all the
data, we conclude that event pile-up is unlikely to have affected the
result.

The contribution of the intrinsic ACIS background cannot be a problem
because it is about 1.5 $\times$ 10$^{-6}$ counts sec$^{-1}$
arcsec$^{-2}$ (Chandra Proposers' Observatory Guide 2002). It is three
order of magnitude lower than that of the trailing events.

\section {\label{sec:discussion} DISCUSSION}

\begin{figure}
\epsscale{1.0}
\plotone{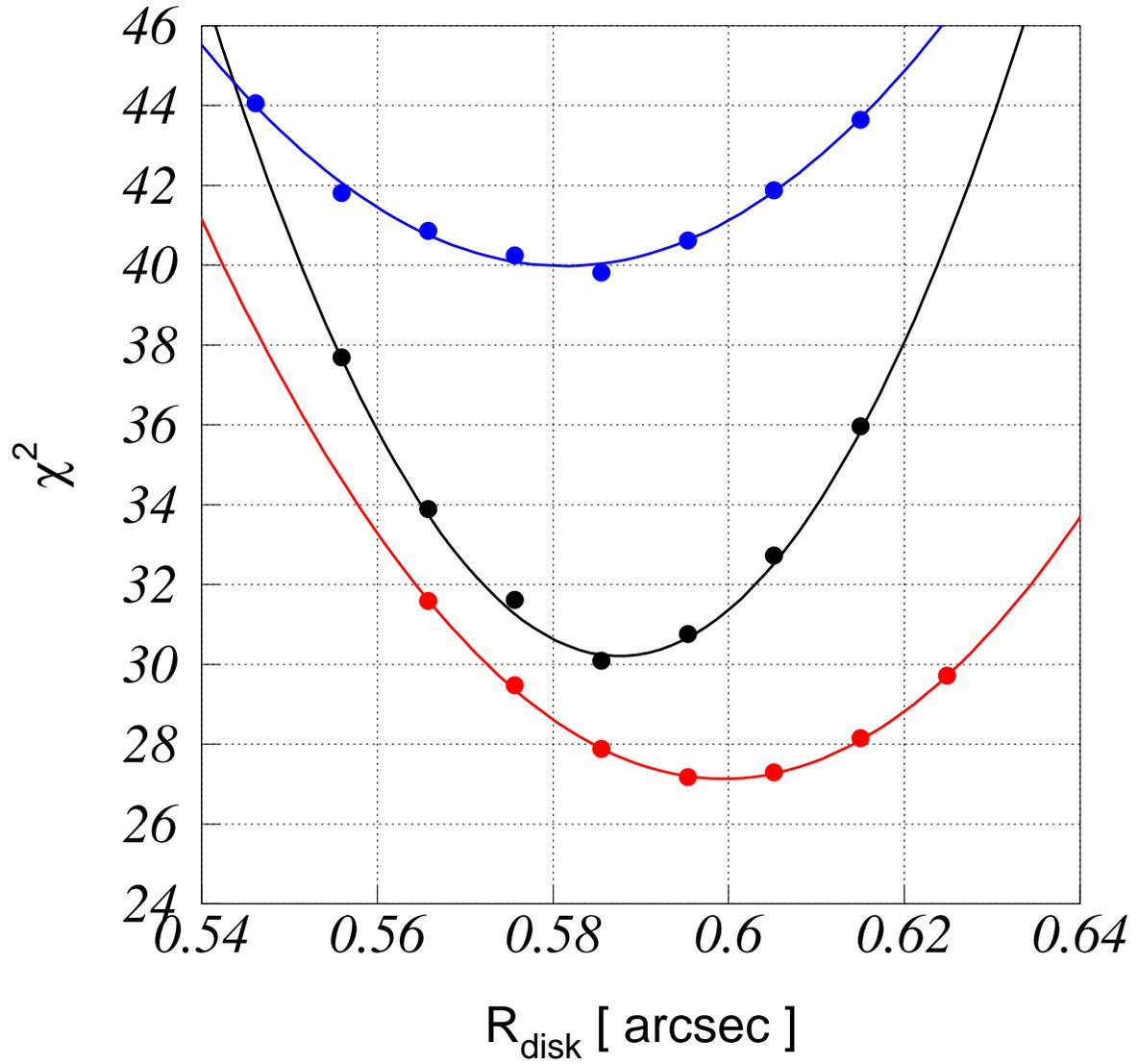}
\caption{
$\chi^{2}$s as a function of $R_{disk}$ (points). Solid curves show
the fits of these points with a quadratic function. Black, red, and
blue colors represent the results for data sets of the entire time
interval, the high flux time interval, and the low flux time interval,
respectively.
\label{fig:chi2}}
\end{figure}

\begin{figure}[]
\epsscale{1.0}
\plotone{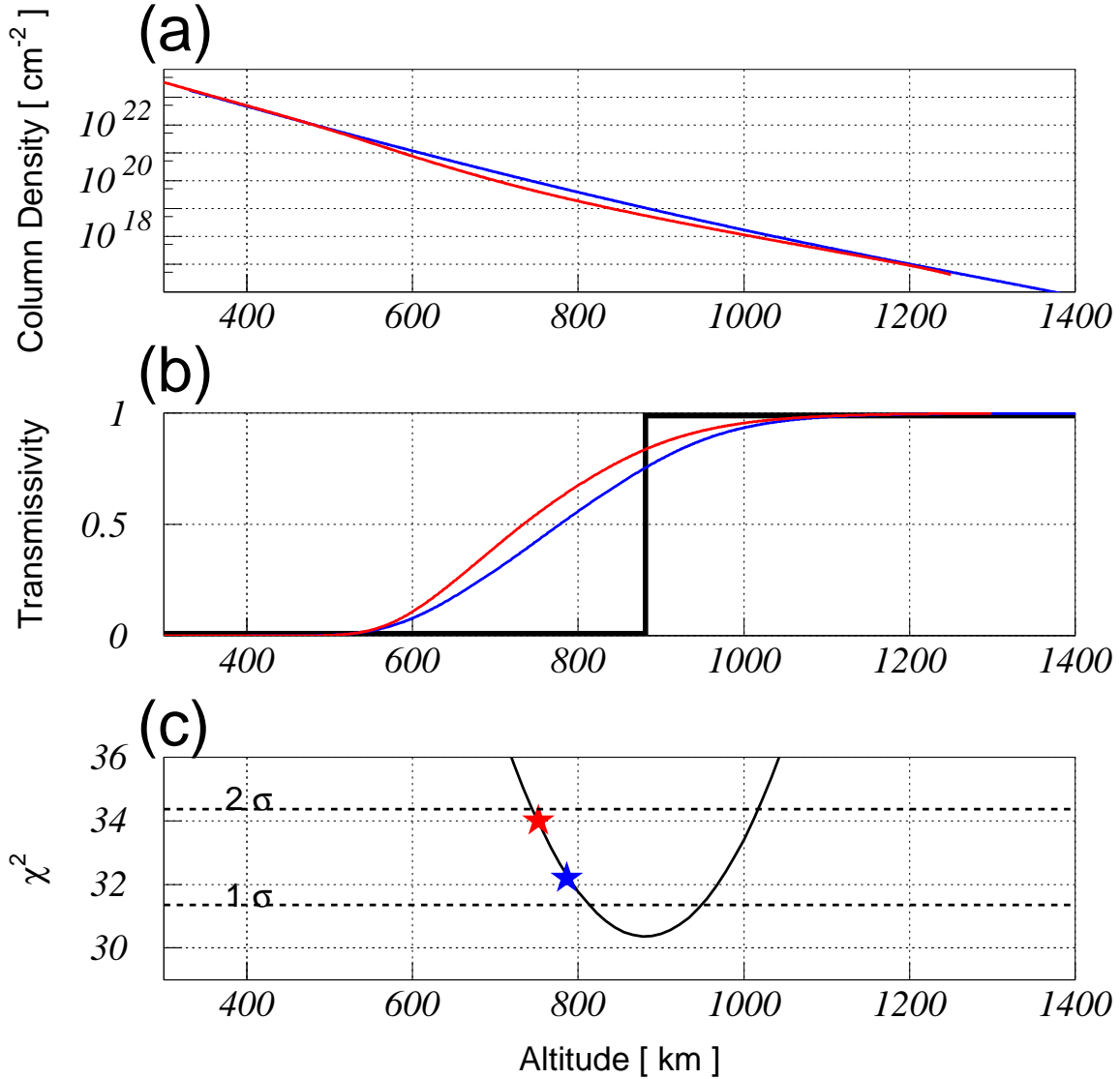}
\caption{ 
(a) Tangential column density along the line-of-sight as a function of
altitude. Red and blue curves were compiled from density profiles
provided by Yelle et al.\ and Vervack et al. (b) X-ray
transmissivities for the observed spectrum as a function of
altitude. The curves are calculated from the tangential column
densities shown in Figure~\ref{fig:comparison}a with the same color
coding. The black line represents the transmissivity assumed in our
best fit disk model. (c) $\chi^{2}$ curve as a function of altitude
obtained for disk model simulations. Red and blue stars are plotted at
($h$, $\chi^{2}$) for simulations with the red and blue transmissivity
curves shown in Figure~\ref{fig:comparison}b, respectively. $h$
represents the characteristic altitude (see text). The horizontal
dashed lines indicate 1 and 2 $\sigma$ confidence levels for the disk
model.
\label{fig:comparison}}
\end{figure}

Taken with the distance of 1.214 $\times$ 10$^{9}$ km to Titan at the
time of this observation, $R_{shadow} = 0.588 \pm 0.011$ arcseconds
gives a thickness for Titan's atmosphere of $880 \pm 60$ km. This
value can be compared with estimates from the models for Titan's
atmosphere which have been constructed based on Voyager~1 observations
at radio, IR, and UV wavelengths. Figure~\ref{fig:comparison}a shows
profiles of the tangential column density (cm$^{-2}$) along the
line-of-sight as a function of altitude (distance above Titan's
surface), which are compiled from density (cm$^{-3}$) profiles
provided by Yelle et al.\ (1997) (red) and Vervack, Sandel, \& Strobel
(manuscript in preparation) (blue). Nitrogen dominates, making up more
than 95\% of the atmospheric constituents, and methane is the second
major component, although proportions slightly differ in the two
models. In those models, the profiles in the altitude range of
500--1000 km, where no data were available, were interpolated between
the data of higher and lower altitudes assuming hydrostatic
equilibrium. The step function we used to determine the atmospheric
thickness is a simplification of the transmission curve of X-rays
through the atmosphere. In reality, X-rays suffer photoelectric
absorption as they pass through the atmosphere, with a transmissivity
determined by the atmospheric composition, the tangential column
density for a given altitude, and photon energy.  We have calculated
transmission curves by using the atmospheric compositions of the above
two models, the tangential column densities in
Figure~\ref{fig:comparison}a, and integrating over the observed Crab
spectrum.  The curves are shown in Figure~\ref{fig:comparison}b. The
step function with a threshold of $R_{shadow}$ is also shown.
Figure~\ref{fig:comparison}c shows a $\chi^{2}$ curve which is
identical with the black curve shown in Figure~\ref{fig:chi2}, but the
definitions of the x-axis are different. We compared the simulation
treating the atmosphere as a solid disk (disk model) with the
simulations based on the transmission curves shown in
Figure~\ref{fig:comparison}b (atmosphere model) as follows. We
calculated $\chi^{2}$ for the atmosphere model simulations and defined
the ``characteristic altitude'' $h$ for those models by

\begin{equation}
h = \int_{0}^{\infty}\, \left( 1-T(h') \right) \,dh',
\end{equation}

\noindent 
where $T(h')$ is transmissivity as a function of altitude, $h'$. We
find that $h$ = 750 and 780 km for the Yelle et al.\ and Vervack et
al.\ models, respectively. The points ($h$, $\chi^{2}$) for those two
atmosphere models fall very close to the $\chi^{2}$ curve for the disk
model, as shown in Figure~\ref{fig:comparison}c.  In other words, our
step-function transmissivity is indistinguishable from a more
realistic transmissivity with $h$ at the threshold of the step in the
simulation.  This fact strongly justifies the approximation of
transmission curve with a step function in our simulation and suggests
that $R_{shadow}$ and $h$ can be directly compared with each
other. Then, our result for the thickness of Titan's atmosphere is
consistent with or slightly larger than those estimated from Voyager~1
observations (1.9 and 1.4 $\sigma$ separation).

Although the statistical significance is not so large, it is still worth
discussing a temporal variation of Titan's upper atmosphere. Distances
between the Sun and Saturn were 1.42 and 1.35 $\times$ 10$^{9}$ km at
the time of Voyager's encounter (November 1980) and our observation
(January 2003), respectively. The solar luminosity is known to be fairly
stable with an uncertainty of 0.3\% regardless of its 11 years active
cycle (Fr\"{o}ehlich \& Lean 1998; Quinn \& Fr\"{o}ehlich
1999). Accordingly, the closer distance to the Sun in our observation
might have resulted in the higher temperature and the larger extent
of the atmosphere.  Although the size of this effect cannot be
ascertained without detailed modeling, the increase in solar flux
incident on Titan may account for some part of the slightly larger value
of Titan's atmosphere measured here. More detailed information will be
obtained by the Cassini/Huygens mission in 2005.

Finally, we note that the spectrum taken from the shadow has no
absorption nor emission line feature and is statistically
indistinguishable from the spectrum taken from the surrounding
region. Those facts indicate that photons penetrating the atmosphere
hardly contributed to the spectrum taken from the shadow; if they did,
excess absorption due to Titan's atmosphere would be seen in the
shadow spectrum. The absence of excess absorption is reasonable
considering that the PSF is much broader (the HPD corresponds to about
5000 km) than the corresponding atmospheric thickness interval of
about 500 km within which the tangential optical depth changes from
unity to zero (see Fig.~\ref{fig:comparison}b).

\acknowledgments

We thank H. Marshall and S. Wolk for supporting observation planning,
R. Yelle and R. Vervack for providing their results, and K. Getman and
E. Feigelson for providing their data. K.\ M.\ and D.\ N.\ B.\ thank
J. Kasting for useful discussions. K.\ M.\ acknowledges the support of
JSPS through the fellowship for research abroad. This work was
supported in part by the NASA through Chandra Award GO3-4002A and was
carried out as a part of ``Ground-based Research Announcement for
Space Utilization'' promoted by the Japan Space Forum.

\end{document}